\documentclass[a4paper, oneside, twocolumn, notitlepage, 10pt]{article}
\usepackage{ecoc}

\usepackage{cite}
\usepackage{amsmath,amssymb,amsfonts}
\usepackage{algorithmic}
\usepackage{graphicx}
\usepackage{textcomp}
\usepackage{xcolor}
\usepackage{gensymb}
\usepackage{makecell}
\usepackage{caption}
\usepackage{sectsty}
\usepackage{titlesec}
\sectionfont{\fontsize{10}{12}\selectfont}
\titlespacing*{\section}
{0pt}    
{12pt}   
{0pt}    
\date{}
\begin{document}

\title{\fontsize{16}{20}\selectfont Silicon Photonic CWDM Filter with Compact Footprint, Low Loss, Flat-Top Transmission and High Yield}

\author{
Qingzhong Deng\textsuperscript{(1)}, 
Alaa Elshazly\textsuperscript{(1, 2)},
Rafal Magdziak\textsuperscript{(1)},
Liesbeth Witters\textsuperscript{(1)},\\
Swetanshu Bipul\textsuperscript{(1)},
Maumita Chakrabarti\textsuperscript{(1)},
Dimitrios Velenis\textsuperscript{(1)},\\
Filippo Ferraro\textsuperscript{(1)},
Huseyin Sar\textsuperscript{(1)},
Peter Verheyen\textsuperscript{(1)},
Philippe Absil\textsuperscript{(1)},\\
Peter Ossieur\textsuperscript{(1)},
Imene Jadli\textsuperscript{(1)},
Joris Van Campenhout\textsuperscript{(1)}
}
\maketitle
\begin{strip}
    \begin{author_descr}

        \textsuperscript{(1)} imec, Kapeldreef 75, 3001 Leuven, Belgium,
        \textcolor{blue}{\uline{qingzhong.deng@imec.be}}

        \textsuperscript{(2)} Photonics Research Group, Department of Information Technology, Ghent University-imec, Ghent, Belgium

    \end{author_descr}
\end{strip}

\begin{strip}
    \begin{ecoc_abstract}
        A novel silicon photonic CWDM filter design is proposed and experimentally demonstrated.
        The design has achieved flat-top transmission across all dies on a wafer, with a device footprint of $48 \times 25~\mathrm{\mu m^2}$, an insertion loss of $0.24 \pm 0.18$ dB, and a channel central wavelength standard deviation of 0.77 nm. ©2026 The Author(s) 
    \end{ecoc_abstract}
\end{strip}

\begin{figure*}[b]
    \centering
    \includegraphics{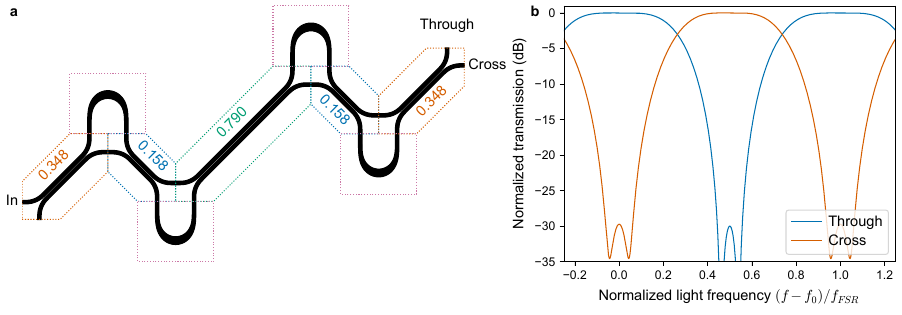}
    \caption{Schematic (a) and theoretical transmission spectra (b) of the proposed MZ4 lattice filter.
    The MZ4 lattice filter comprises four identical phase shifters interconnected by five directional couplers with power cross-coupling ratios as indicated in the figure.
    \label{fig_mz4_theoretical_spectra}
    }
\end{figure*}

\begin{figure*}[t]
    \centering
    \includegraphics{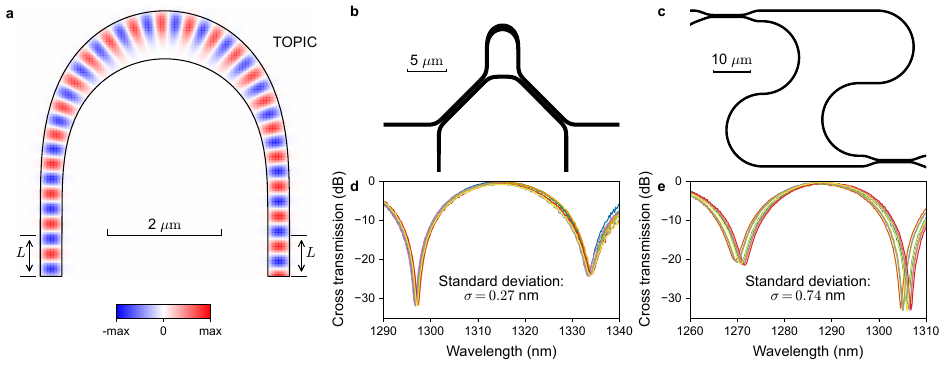}
    \caption{(a) Optical field evolution (H along the thickness direction) in the TOPIC bend phase shifter.
    The simulation was performed using COMSOL with a 3D full-vector finite-element method, and the plotted field evolution is sliced at the waveguide center along the thickness direction at light wavelength of 1310 nm.
    The TOPIC bend phase shifter has a radius of 2 $\mathrm{\mu m}$, port waveguide width of 0.38 $\mathrm{\mu m}$, and transition angles of $\theta_{p,i}=42.5^{\circ}$ and $\theta_{p,o}=65^{\circ}$ for the inner and outer boundaries, respectively.
    Two identical straight waveguide segments, with length $L$, connected at the input and output ports enable phase delay adjustment for a specific FSR.
    Schematics (b, c) and measured spectra (d, e) of the proposed compact MZI with TOPIC bends (b, d) and the conventional MZI with circular bends (c, e).
    The spectra are measured from 9 repeated designs placed close to each other in the same die.
    \label{fig_mz1_phase_error}
    }
\end{figure*}

\begin{figure*}[b]
    \centering
    \includegraphics{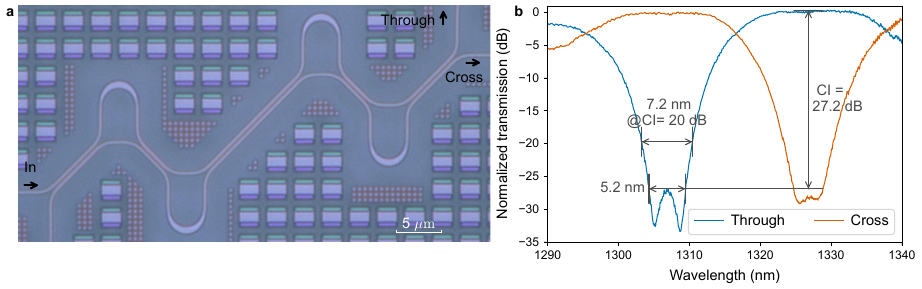}
    \caption{Microscope image (a) and measured transmission spectra (b) of the proposed MZ4 lattice filter.
    The TOPIC bend phase shifters use the same structural parameters as described in Fig.~\ref{fig_mz1_phase_error} with $L=0.72\ \mathrm{\mu m}$.
    This device is fabricated on imec's 300 mm silicon photonics platform (isipp300) using a standard process with a 220 nm silicon thickness and a 2 $\mathrm{\mu m}$ buried oxide layer.
    \label{fig_mz4_measured_spectra}
    }
\end{figure*}

\begin{figure*}[t]
    \centering
    \includegraphics{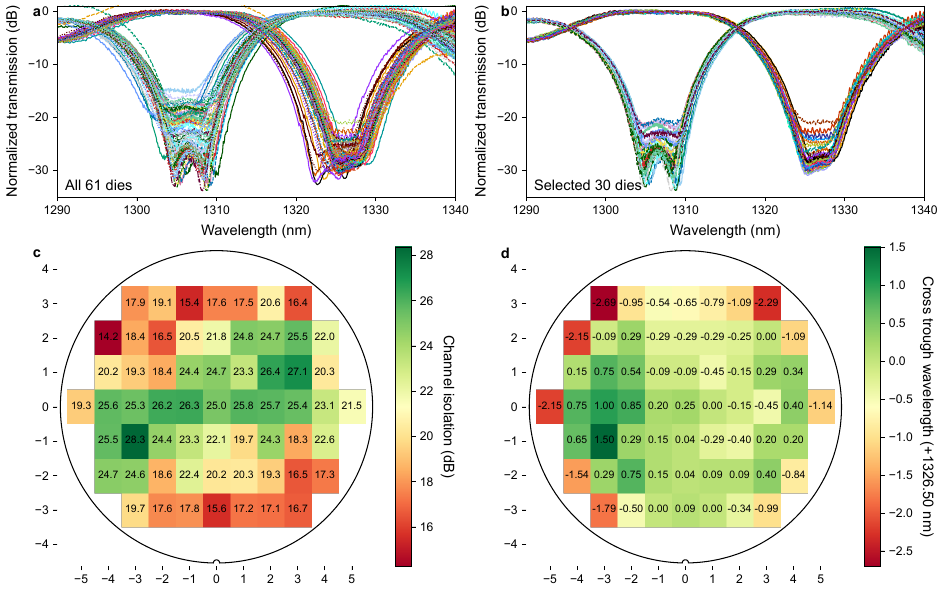}
    \caption{Measured MZ4 transmission spectra from all the 61 dies (a) and 30 selected dies (b) in one wafer. Wafer mapping of the measured channel isolation (c) and Cross trough wavelength (d).
    \label{fig_mz4_wafer_mapping}
    }
\end{figure*}
\section*{Introduction}
Silicon photonics enables high-capacity optical interconnects through wavelength division multiplexing (WDM).
Coarse WDM (CWDM) with 20 nm channel spacing is widely used in short-reach systems. 
In practical applications, CWDM filters are expected to work without power consumption for tuning or stabilization, which requires two critical features: flat-top transmission and high fabrication tolerance.
Cascaded Mach-Zehnder (MZ) lattice filters are commonly used to achieve flat-top transmission~\cite{CascadedMachZehnder_OE2013a}, but their sensitivity to fabrication variations limits yield.
To improve the fabrication tolerance, one or more wide waveguide segments are introduced into the phase shifters of the MZ filters~\cite{SiliconWireOptical_AO2018a,FabricationTolerantO_JOLT2024a,FabricationTolerantCwdm_JOLT2021a,LowChannelCrosstalk_COLAEC2023a,RobustDesignSilicon_IJOEASTICAS2026a,FabricationTolerantCwdm_JOLT2025a,VariationAwareLayout_2023a,LowCrosstalkFabrication_OE2021a,LayoutAwareVariability_IJOSTIQE2019a,MaximizingFabricationThermal_IPTL2015a}.
However, flat-top transmission have been reported only on a few (2~\cite{SiliconWireOptical_AO2018a}, 
5~\cite{FabricationTolerantO_JOLT2024a},
9~\cite{FabricationTolerantCwdm_JOLT2021a},
20~\cite{LowChannelCrosstalk_COLAEC2023a}
) selected dies when the Silicon CWDM filters are massively produced at wafer scale.
The low yield is primarily attributed to random phase errors induced by fabrication variations.
The accumulation of random phase errors increases with phase shifter length.
Although the aforementioned fabrication-tolerant designs reduce spectral shifts when both MZI arms experience similar waveguide width and thickness variations, they cannot mitigate the impact of random phase errors.
Furthermore, these designs introduce additional random phase errors because tapers are required to connect waveguide segments of different widths, which increases the total phase shifter length.
Moreover, the widely used flat-top MZ design~\cite{CascadedMachZehnder_OE2013a} necessitates doubling the phase shifter length relative to the length defined by the required free spectral range (FSR), thereby further increasing random phase errors.

In this paper, we demonstrate a silicon photonic CWDM filter with three distinctive features compared to state-of-the-art designs: 
(1) a novel MZ lattice filter that achieves flat-top transmission with phase shifter length defined solely by the required FSR;
(2) a compact layout design that minimizes routing lengths in the phase shifters~\cite{SiliconRingBased_2024a}; and
(3) gradually widening phase shifters using a third-order polynomial interconnected circular (TOPIC) bend~\cite{LowLossLow_LPR2024a} that eliminates the need for tapers.
These features collectively reduce waveguide length in the phase shifters, thereby minimizing the accumulation of random phase errors and propagation loss, which results in high yield across all dies on the wafer with flat-top transmission, compact footprint, and low insertion loss.

\section*{MZ4 Lattice Filter}
Fig.~\ref{fig_mz4_theoretical_spectra}a shows the schematic of the proposed CWDM filter, referred to as the MZ4 lattice filter, which comprises four identical phase shifters interconnected by five directional couplers.
When the power cross-coupling ratios are set to 0.348, 0.158, 0.790, 0.158, and 0.348 for the directional couplers respectively, the filter achieves flat-top transmission at both the Through and Cross ports, as shown in Fig.~\ref{fig_mz4_theoretical_spectra}b.
To minimize the phase shifter length, we employ a compact layout design that reduces routing lengths in the phase shifters~\cite{SiliconRingBased_2024a}.
The directional couplers feature a specialized design with output ports oriented at 90$^\circ$ to each other, and the same configuration is applied to the input ports.
When the five directional couplers are connected as shown in Fig.~\ref{fig_mz4_theoretical_spectra}a, the relative phase shift between the two arms of each lattice stage can be implemented using a single 180$^\circ$ bend.
A 180$^\circ$ TOPIC bend is used to implement the phase shifters~\cite{LowLossLow_LPR2024a}.
As shown in Fig.~\ref{fig_mz1_phase_error}(a), the TOPIC bend with a 2 $\mathrm{\mu m}$ radius can gradually widen the waveguide from 0.38 $\mathrm{\mu m}$ to 0.84 $\mathrm{\mu m}$ while maintaining single-mode propagation as a whispering gallery mode.

Traditional MZ filters contain extensive common routing waveguide segments in both arms (Fig.~\ref{fig_mz1_phase_error}c), which do not contribute to the FSR of the MZ filter but increase the accumulation of random phase errors.
In contrast, the proposed compact MZ design minimizes common routing waveguide segments in both arms, as shown in Fig.~\ref{fig_mz1_phase_error}b.
When both designs are configured for the same FSR ($\sim$6.4 THz) and fabricated using identical processes, the compact MZ design exhibits significantly reduced phase errors, with the peak wavelength standard deviation decreased to 0.27 nm (Fig.~\ref{fig_mz1_phase_error}d) from 0.74 nm (Fig.~\ref{fig_mz1_phase_error}e).

Fig.~\ref{fig_mz4_measured_spectra}a shows the microscope image of the fabricated MZ4 lattice filter, which has a compact footprint of 48$\times$25 $\mathrm{\mu m^2}$.
The measured transmission spectra in Fig.~\ref{fig_mz4_measured_spectra}b show that the MZ4 lattice filter achieves flat-top transmission with a channel spacing of 19.8 nm, a 0.5 dB bandwidth of 12.5 nm, an insertion loss of almost 0 dB, a channel isolation (CI) of 27.2 dB over 5.2 nm, and a passing bandwidth of 7.2 nm at CI = 20 dB.
This device is characterized over all 61 dies in one wafer, with wafer mapping of the measured channel isolation and cross trough wavelength plotted in Fig.~\ref{fig_mz4_wafer_mapping}c and d respectively.
As shown in Fig.~\ref{fig_mz4_wafer_mapping}a, all 61 dies exhibit flat-top transmission with a CI $\geq$ 14.2 dB.
The standard deviation of the Cross though wavelength is 0.77 nm only.
The worst channel insertion loss is 0.24 $\pm$ 0.18 dB.
The channel spacing is 19.89 $\pm$ 0.08 nm.
The 0.5 dB bandwidth is 11.4 $\pm$ 1.3 nm with a CI of 21.5 $\pm$ 3.7 dB for the Through spectra, 
and the 0.5 dB bandwidth is 11.1 $\pm$ 1.2 nm with a CI of 27.1 $\pm$ 2.0 dB for the Cross spectra.
Fig.~\ref{fig_mz4_wafer_mapping}b shows the measured transmission spectra from 30 selected dies with CI $\geq$ 20 dB, where 
the standard deviation of the Cross though wavelength is 0.32 nm,
the worst channel insertion loss is 0.19 $\pm$ 0.15 dB, 
the channel spacing is 19.90 $\pm$ 0.06 nm,
the 0.5 dB bandwidth is 11.6 $\pm$ 0.9 nm with a CI of 24.1 $\pm$ 2.0 dB for the Through spectra, 
and the 0.5 dB bandwidth is 11.3 $\pm$ 1.0 nm with a CI of 26.7 $\pm$ 2.3 dB for the Cross spectra.

\section*{Conclusion}
In conclusion, we have demonstrated a silicon photonic CWDM filter that achieves flat-top transmission across all dies on a wafer.
The proposed MZ4 lattice filter, combined with compact layout design and TOPIC bends, reduces phase shifter length and minimizes random phase errors induced by fabrication variations.
Measurements from 61 dies demonstrate that all devices exhibit flat-top transmission with channel isolation exceeding 14.2 dB and a central wavelength standard deviation of only 0.77 nm, demonstrating unprecedented yield for silicon photonic CWDM filters.
The filter achieves a compact footprint of $48 \times 25~\mathrm{\mu m^2}$, a worst channel insertion loss of $0.24 \pm 0.18$ dB, and excellent spectral performance, making it well-suited for practical deployment in high-capacity optical interconnect systems.

\clearpage
\section*{Acknowledgements}
This work was supported by imec's industry-affiliation R\&D program “Optical I/O”.

\bibliographystyle{IEEEtran}
\bibliography{flat_top_cwdm_ecoc2026}
\end{document}